# Non-collinear rotational Doppler effect

Aleksandr Bekshaev [a] and Andrey Popov [b]

[a] I.I. Mechnikov National University, Dvoryanska 2, 65026, Odessa, Ukraine
[b] Research Institute of Physics, I.I. Mechnikov National University, Pastera, 27, 65026, Odessa, Ukraine

**ABSTRACT**

The frequency shift of a helical light beam experiencing the rotation near the axis differing from its own axis (conical evolution) is studied theoretically. Both the energy and the kinematic approaches lead to a paradoxical conclusion that after a whole cycle of the system rotation the beam does not return to its initial state. Another paradox is manifested in the peculiar behavior of the beam transverse pattern rotation at different geometric parameters of the evolving system. A fundamental role of the detecting system motion is substantiated. The special "natural" observer's motion is found for which both paradoxes are eliminated. Relations of the described facts with the Hannay's geometric phase concept are discussed.

## 1. INTRODUCTION: DOPPLER EFFECT AND ITS PHYSICAL INTERPRETATION

As it is well known [1], the Doppler effect consists in the fact that the observable wave frequency depends on the relative motion of the wave source and the observer. For example, if a monochromatic beam with frequency $\omega$ and wave number $k$ propagates along axis *z*, in the laboratory frame its field distribution contains the factor

$$h(z,t) = \exp\left[i(kz - \omega t)\right]. \tag{1}$$

If an observer moves along the same axis, it corresponds to the coordinate transformation which in the non-relativistic case reads

$$z = z' - vt$$

and, in the observer's frame, factor (1) obtains the form $h(z',t) = \exp\left[i(kz' - \omega't)\right]$ where

$$\omega' - \omega = \Delta\omega = kv \tag{2}$$

is an expression of the common translational Doppler frequency shift.

In the last time, the considerable attention is paid to a new phenomenon of the rotational Doppler effect (RDE) [2–4] which is a peculiar property of waves with helical structure, for example, circularly polarized light [3–6] or circular Laguerre-Gaussian (LG) modes [7–13] (see Fig. 1). The helical structure of such beams is described by the analogue of factor (1)

$$h(z,\phi,t) = \exp\left[i(kz + l\phi - \omega t)\right]$$

where $\phi$ is the azimuth angle within the beam cross-section and *l* is the helicoid winding number (azimuthal index of the LG mode). For the RDE, not the translation along the beam axis but the rotation around it is important. The observer's plane of analysis is always orthogo-nal to axis $z$ and within this plane a certain reference axis exists (e.g., the polar axis with respect to which azimuth angles are determined). In this frame, mutual rotation of the beam and the observer with angular velocity $\Omega$ can be expressed by transform $\phi = \phi' - \Omega t$ whence the so called rotational frequency shift

$$\Delta\omega = l\Omega \qquad (3)$$

immediately follows. Note that in contrast to the usual Doppler shift (2), the RDE shift does not depend on the beam wavelength.

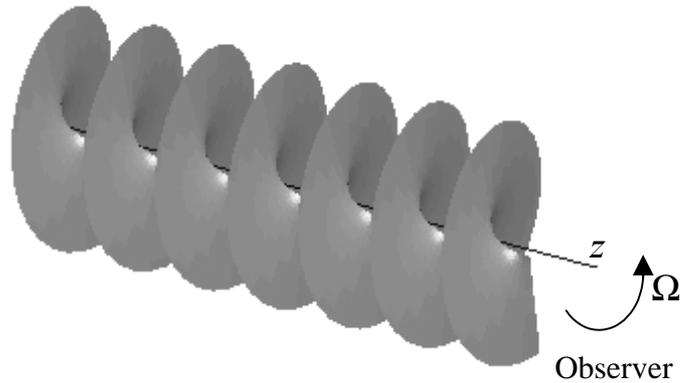

Fig. 1. Helical beam structure.

The Doppler frequency shift can also be treated on the base of energy exchange between the beam and the optical elements [3, 6, 7] in which the "interactive aspect" [3] of the Doppler effect is manifested. This energy exchange occurs due to the mechanical properties of the light. For example, a helical beam carries the mechanical angular momentum (AM) [3, 13–15]. If such a beam interacts with an optical system which changes the AM, it leads to a ponderomotive torque applied to the optical system as well as to the optical field [3, 16]. To be certain, consider a usual scheme of the RDE observation where the rotation with respect to a fixed observer is imparted to a beam by means of a rotating Dove prism [8, 9] or a three-mirror system [16] (Fig. 2).

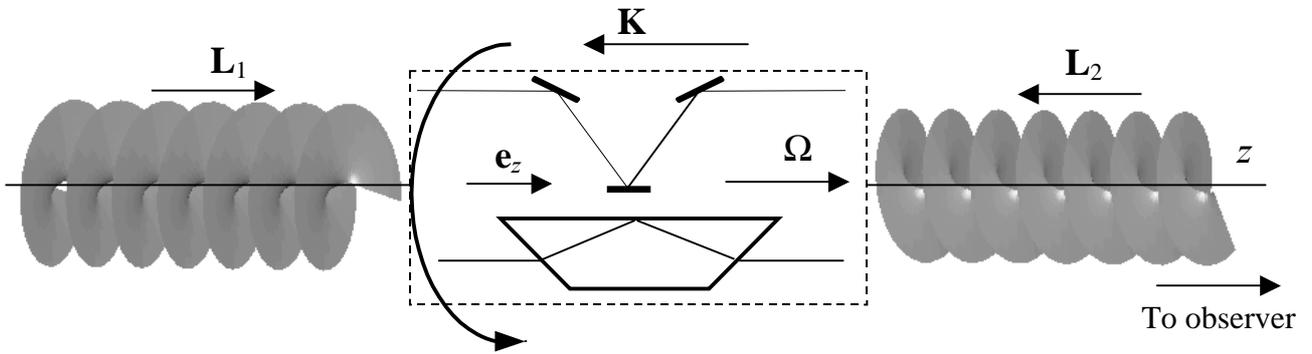

Fig. 2. Schematic of the collinear RDE. The dashed box comprises two variants of the rotating optical system; their transverse displacements are conventional (normally their optical axes coincide with axis $z$), $\mathbf{e}_z$ is a unit vector of axis $z$. Other notations are explained in the text.

Here the input beam is represented by an LG mode with positive $l$ and the AM linear density

$$\mathbf{L}_1 = Nl\hbar\mathbf{e}_z, \qquad (4)$$

or $l\hbar$ per photon [14]; $N$ is the linear density of photons so that every second of time $Nc$ photons pass through the system cross section ($c$ is the light velocity). The optical system of Fig. 2 reverses the beam handedness and, of course, its AM: $\mathbf{L}_2 = -Nl\hbar\mathbf{e}_z$, applying to the optical field the mechanical torque

$$\mathbf{K} = c(\mathbf{L}_2 - \mathbf{L}_1) \text{ with absolute value } |\mathbf{K}| = 2Nl\hbar c. \qquad (5)$$

Simple geometrical considerations show [8] that when the optical system rotates with the angular velocity $\Omega$, the output beam rotates with double velocity and since its azimuthal index is $-l$, Eq. (3) gives for this case

$$\Delta\omega = -2l\Omega. \qquad (6)$$

On the other hand, in this rotation, the torque (5) produces the mechanical work per second $(\mathbf{K}, \mathbf{\Omega}) = -2Nl\hbar c\Omega$ (the usual notation of the scalar product is used). This value equals to the variation of energy of the beam sector passing the system per second and containing $Nc$ photons so that each photon obtains the energy gain

$$\Delta(\hbar\omega) = -2l\hbar\Omega$$

which corresponds to the beam frequency change exactly coinciding with the outcome of geometrical analysis (6).

## 2. CONICAL BEAM EVOLUTION

In the usual arrangement for the RDE observation (e.g., one shown in Fig. 2), axis of the beam under examination and the axis of rotation coincides; this case can be called a "collinear RDE". But a lot of physical situations may exist where the axis of the beam rotation differs from its own axis; we shall term them as a "non-collinear RDE". An example of this situation is provided by Fig. 3. where the tilted mirror is obliquely attached to a rotary shaft After reflection from the tilted mirror, the output beam axis moves over the conical surface; therefore, in this scheme the so-called conical beam evolution takes place.

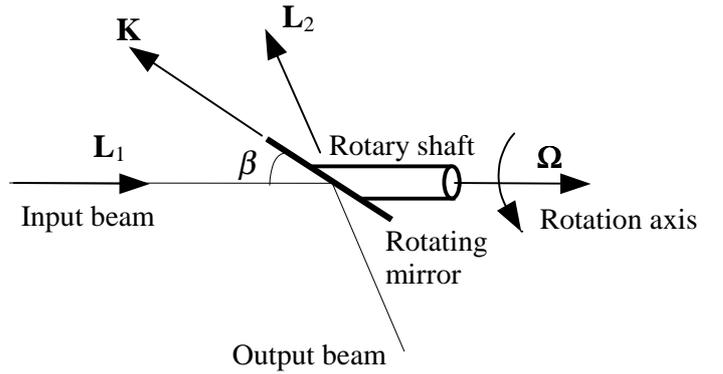

Unlike the case of collinear RDE considered in previous section, we start the non-collinear RDE analysis with the energy approach. Let the input beam be again an LG mode with positive $l$ and AM $\mathbf{L}_1$ given by (4); the output beam possesses the azimuthal index $-l$ and the same absolute value of AM $\mathbf{L}_2$ but its direction changes as shown in Fig. 3. Operating as before, we can calculate the torque applied to the beam; its direction is shown in Fig. 3 and absolute value is

$$|\mathbf{K}| = 2Lc\cos\beta = 2Nl\hbar c\cos\beta;$$

When the shaft rotates, this torque changes the output beam energy, and the same deduction as in previous section leads to the following expression for the Frequency shift of the output beam

Fig. 3. Scheme of the non-collinear RDE observation in the course of conical evolution of the output beam

$$\Delta\omega = -2l\,\Omega\cos^2\beta. \tag{7}$$

Let us analyze this result. Obviously, it is equivalent to a certain shift of the output beam phase obtained during the mirror rotation through a certain angle $\theta = \Omega t$. Evoking the phase factor (1), we can present it temporal part in the form

$$\exp(-i\omega' t) = \exp(-i\omega t)\exp(-i\Delta\omega t) = \exp(-i\omega t)\exp(i\Delta\varphi),$$

where

$$\Delta\varphi = 2l\Omega t\cos^2\beta = 2l\theta\cos^2\beta. \tag{8}$$

Note that this value exactly corresponds to the phase shift calculated for the same case on the base of the adiabatic invariance principle [16].

Simple formula (8) leads to an unexpected consequence. Being applied to the case of a whole cycle of the mirror revolution ($\theta = 2\pi$), it gives

$$\Delta\varphi = 4\pi l\cos^2\beta. \tag{9}$$

So, after a whole cycle of the system revolution, when its configuration returns to the initial state, the output beam phase does not return to its initial value; it experiences the non-integer (in units of $2\pi$) phase shift.

This conclusion is even more paradoxical than it seems. The phase of a helical light beam can be visualized, e.g., due to superposition with a Gaussian beam, which produces the system of $|l|$ so-called off-axial optical vortices [11, 12] (a spe-cific beam pattern with the dark point in the beam spot periphery). The azimuthal displacement $\Delta\phi$ of these dark points is determined by the phase of the helical component as $\Delta\phi = \Delta\varphi / l$ [11, 12] (see the example for $l = 1$ in Fig. 4). Therefore, the non-integer phase shift means that after the mirror performs a whole revolution, the output beam pattern is not restored but will be turned by some angle depending on the mirror tilt and following from Eq. (9)

$$|\Delta\phi| = 4\pi\cos^2\beta \qquad (10)$$

which sharply contradicts to our daily experience.

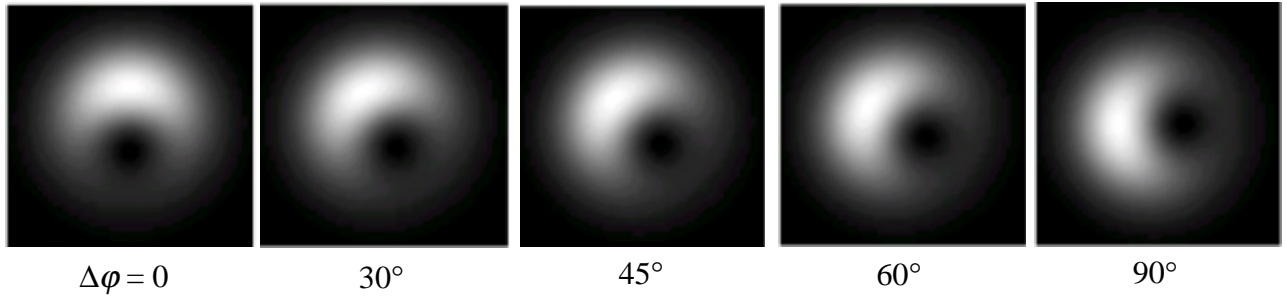

$\Delta\varphi = 0$  30°  45°  60°  90°

Fig. 4. Transverse beam pattern of the superposition of a Gaussian and $LG_{01}$ modes; each image is labeled by the relative phase of the LG component.

To see another view of the situation, consider the beam spatial transformation in the course of the conical evolution in more detail. Imagine that the input beam carries some image (Fig. 5). Then the output beam image demonstrates two qualitatively different modes of behavior depending on the mirror tilt angle $\beta$. In case of "forward reflection" ($\beta < 45°$) an output image performs two full cycles of revolution when the mirror makes one cycle (see Fig. 5). At $\beta > 45°$, when the output beam goes to the back hemisphere, the output image does not rotate at all!

The first configuration can be continuously transformed into the second one by changing the mirror tilt angle. But at any tilt angle, one can observe only an integer number of cycles of the image rotation. The continuous transition between the "forward reflection" and "back reflection" cases in the image behavior seems impossible.

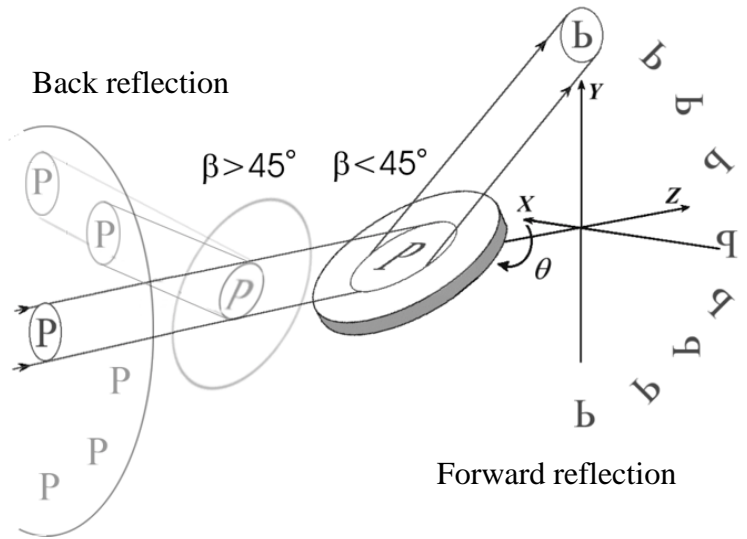

Fig. 5. Image transformation at the conical beam evolution

Therefore, a simple analysis of the helical beam conical evolution leads to conclusions which seem, at first sight, paradoxical:

(I) Energy arguments predict that such a beam acquires a non-integer phase shift after a whole cycle of the tilted mirror revolution. Contradictory to experiment, this means that the spot pattern of an off-axial vortex generally does not restore its initial orientation after the system returns to its initial state.
(II) The output image rotation demonstrates qualitatively different modes of behavior, which apparently change by jumps when the mirror tilt angle changes continuously.

For the explanation, ideas of quantization of the photon angular momentum projection on the rotation axis were engaged [17]. To our opinion, there is no need in so extravagant construction, and the whole situation can be understood from the classical point of view.

## 3. CONICAL BEAM EVOLUTION: KINEMATIC ANALYSIS

Consider immediately the geometric picture of the conical beam evolution with the help of Fig. 6 which, in its main parts, is taken from Ref. [16], and Fig. 7. The figures correspond to the conical evolution arrangement of Fig. 3.

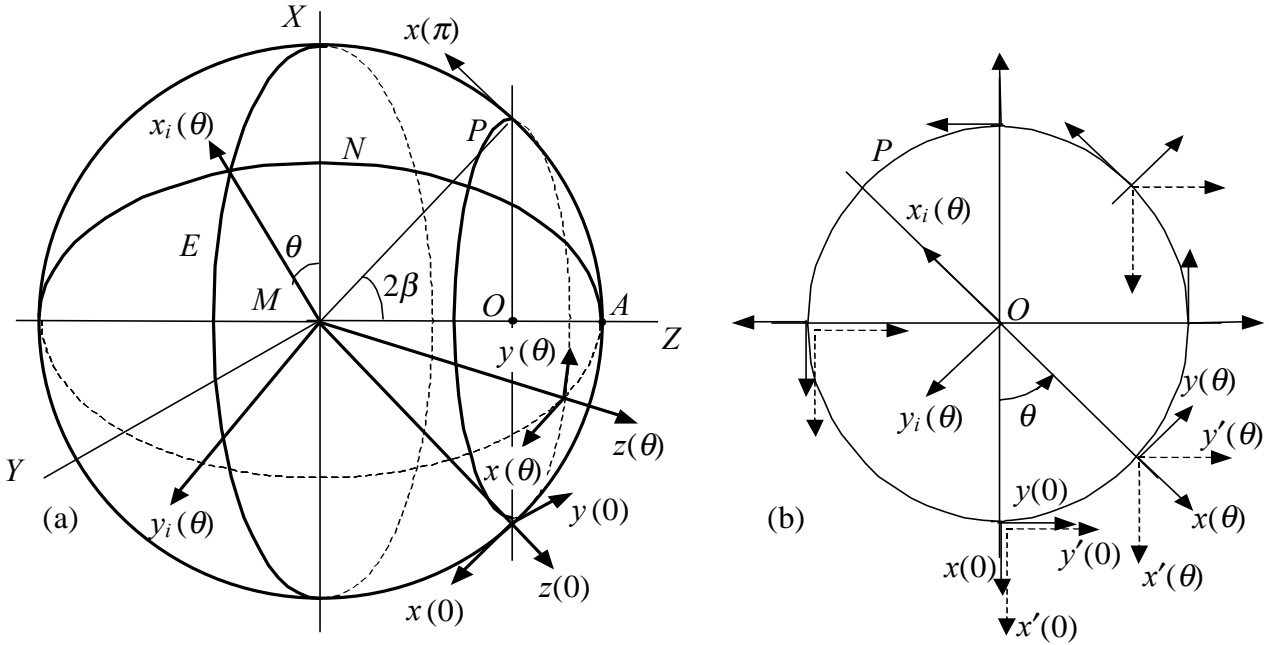

Fig. 6. Disposition of coordinate systems in the problem of conical beam evolution:
(a) general view in perspective (dashed segments of "equator" $E$, "meridian" $N$ and "parallel" $P$ are "covered" by the sphere), (b) view of parallel $P$ from the positive semi-axis $Z$.

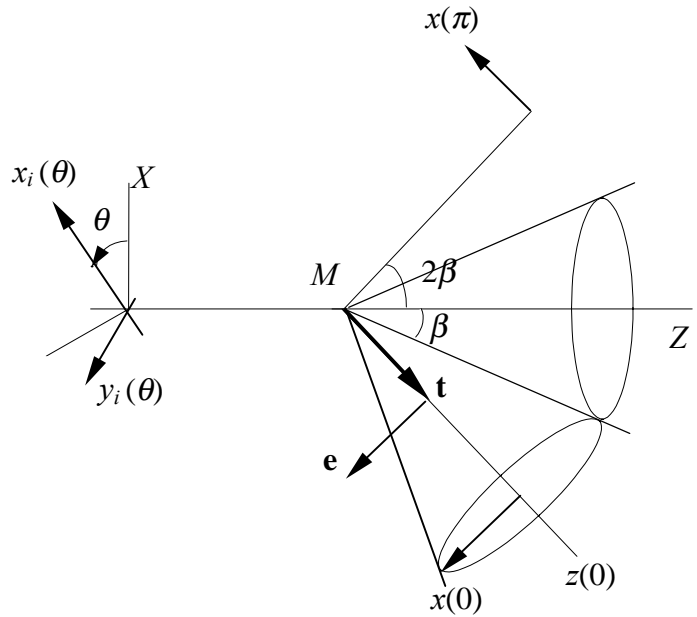

Fig. 7. Kinematic model of the conical evolution performed by the system of Fig. 3 (explanations in text).

The analysis is suitable with the help of auxiliary "earthlike" sphere centered at the point $M$ of intersection between the tilted mirror and the axis of rotation $Z$. Point $A$ where axis $Z$ crosses the sphere is an analogue of the Earth pole. We introduce several coordinate systems pertinent to the problem. Coordinate system $XYZ$ is a laboratory frame; its origin coincides with point $M$ and the coordinate plane $Z = 0$ coincides with the sphere equatorial plane $E$. The input beam propagates along the axis $Z$; when the tilted mirror rotates, the output beam axis $z(\theta)$ moves along the "parallel" with latitude $\pi/2 - 2\beta$ (circumference $P$ with center $O$). Axes $x_i(\theta)$, $y_i(\theta)$ lie in plane $Z = 0$ but are connected with the mirror so that axis $y_i(\theta)$ is parallel to it and $x_i(\theta)$ is directed oppositely to the current projection of the output beam axis. The system $xyz(\theta)$ is connected to the output beam: axis $z(\theta)$ always coincides with the beam's own axis and axes $x(\theta), y(\theta)$ belong to its cross section. Their current directions agree with the mirror position so that $x(\theta)$ belongs to the plane formed by axes $x_i(\theta)$ and $Z$ (meridian plane $N$ of the sphere) and $y(\theta)$ is directed along the parallel $P$ (see Fig. 6a). In other words, frame $x(\theta), y(\theta)$ defines local coordinates within the "earthlike" sphere surface; some possible projections of these axes are shown as radial and tangent arrows along the circumference in Fig. 6b. We will also need another coordinate system $x'y'z(\theta)$, third axis of which coincides with axis $z(\theta)$ and the first two axes preserve the directions that are "as close as possible" to the directions of the unmovable axes $X, Y$ (projections of these axes are shown in Fig. 6b by dashed lines). This frame is formed from the unmovable frame $XYZ$ by "turning" in the shortest way in order that the output beam cross-section coincide with plane $z = 0$, plus reversal of axes $x', y'$, or from the system $xyz(\theta)$ by turn through angle $\theta$ oppositely to the beam evolution direction. We will assume the origins of all three coordinate systems to situate in point $M$ (in Figs. 6, 7 they are displaced for the convenience). Coordinates of a point in different frames are related by the transformation rules that are given in Appendix 1.

In this representation, the tilted mirror "rolls" without slip over the conical surface with axis $Z$ and the cone generator angle $\beta$ (Fig. 7). The output beam appears to be "trapped" within another identical cone which rolls without slip over the side of the first cone. Both cones have the common apex in point $M$. The motion of the unit vector $\mathbf{t}$ of the moving cone axis in the coordinate frame $XYZ$ is described by equation

$$\mathbf{t}(\theta) = \begin{pmatrix} -\sin 2\beta \cos\theta \\ -\sin 2\beta \sin\theta \\ \cos 2\beta \end{pmatrix} \qquad (11)$$

To trace the beam transformation upon the mirror rotation, it is sufficient to consider the motion of arbitrary vector $\mathbf{R}$, rigidly bounded with the moving cone; let it be the image of the unit vector of axis $X$, which in the system $xyz(\theta)$ moves in accord with the law $x = \cos\theta$, $y = \sin\theta$, $z = 0$ with the angular velocity $\Omega = \dot{\theta}$ (upper point denotes the time derivative); in the frame $x'y'z(\theta)$ this velocity will amount

$$\Omega' = 2\Omega. \qquad (12)$$

Coordinates of this vector in the laboratory frame are determined by Eq. (A.1) with the transformation matrix (A.2). In the outcome, we obtain

$$\mathbf{R}(\theta) = \begin{pmatrix} \sin^2\beta - \cos^2\beta \cos 2\theta \\ -\cos^2\beta \sin 2\theta \\ -\sin 2\beta \cos\theta \end{pmatrix}. \qquad (13)$$

These equations give the exhaustive kinematical description of the output beam behavior. In particular, they determine the velocity vector of motion (13)

$$\mathbf{V}(\theta) = \dot{\mathbf{R}}(\theta) = \Omega \begin{pmatrix} 2\cos^2\beta \sin 2\theta \\ -2\cos^2\beta \cos 2\theta \\ \sin 2\beta \sin\theta \end{pmatrix}. \qquad (14)$$

Since $\mathbf{V}(\theta)$ is a velocity of the end of the unit vector trapped within the output beam, it can easily be related to the angular velocity of rotation of the transverse beam structure (or an image carried by the beam like that of Fig. 5) around its mobile axis (let us call this motion "proper" beam rotation). To this purpose, find the vector (14) component lying within the beam cross section, i.e. normal to the beam axis unit vector (11):

$$\mathbf{V}_\perp(\theta) = \mathbf{V}(\theta) - \mathbf{t}(\mathbf{V}(\theta), \mathbf{t}) = 2\Omega \cos^2\beta \begin{pmatrix} \cos^2\beta \sin 2\theta \\ -\sin^2\beta - \cos^2\beta \cos 2\theta \\ \sin 2\beta \sin\theta \end{pmatrix}. \qquad (15)$$

Then the angular velocity of the "proper" rotation will be determined by the relation (square brackets mean the vector product)

$$\mathbf{\Omega}_1 = [\mathbf{R}(\theta), \mathbf{V}_\perp(\theta)] = 2\Omega \cos^2\beta \cdot \mathbf{t}. \qquad (16)$$

As should be expected, the absolute value of the angular velocity does not depend on the mirror turn $\theta$ and its direction coincides with the moving beam axis (11). Herewith, the angular velocity of the beam rotation is not multiple of the mirror rotation velocity which seems to cause the output beam's not returning to its initial position after the whole cycle of the mirror revolution around axis *Z*. In compliance with (16), it will turn through the angle $2 \cdot 2\pi \cos^2\beta$.

This conclusion is rather remarkable because it gives the kinematic confirmation to the result of the energy-based consideration (10). However, it does not solve the paradox (I) formulated at the end of Sec. 2 but makes it still stronger: now, two different ways of deduction lead to the same prediction conflicting to reality. How to agree the non-integer beam revolution, expressed by (10), with observable image reproduction, still remains a problem.

### 4. MOTION OF THE REFERENCE SYSTEM

To solve this puzzle, let us carefully consider main conditions of the non-collinear RDE observation. The matter is that the observed phase of a helical beam depends on the relative disposition of the beam and the detector system. A helical beam phase $\varphi$ (and frequency $\omega = d\varphi/dt$) has no absolute meaning; for the same beam, different observers may see different phases. In this connection, the questions appear:
- Which reference system ("observer") is correct, or "natural", for this case?
- To which reference system the above energy and kinematic considerations relate?

Note, by the way, that analogous questions are implicitly present in the usual case of the collinear RDE. But there they are answered trivially: since the output beam axis is fixed, the natural choice of a fixed observer appears to be correct. In the non-collinear case, an observer, whose plane of analysis should be always orthogonal to the moving output beam axis (see Sec. 1), is obliged to move. Simultaneously, the reference axis in this plane, which provides a benchmark for the beam rotation, cannot stay unmovable and the question about the character of this motion becomes substantial.

Let the reference axis position is determined by the vector **e** lying within the output beam cross section. In the course of the conical evolution this vector remains constant, from the certain "intrinsic" point of view, but moves with respect to the laboratory frame. Then, the apparent reproduction of the beam transverse orientation after a full system revolution can be explained as a result of superposition of the "non-integer" proper" beam revolution (10) and of the angular motion of the reference system itself.

Now, try to determine the "natural" motion of vector **e** upon the tilted mirror rotation. We will proceed from the following assumptions:

(a) vector **e** is always orthogonal to **t**, so that the condition $(\mathbf{e}, \mathbf{t}) \equiv 0$ and its consequence $(\dot{\mathbf{e}}, \mathbf{t}) = -(\mathbf{e}, \dot{\mathbf{t}})$ hold;

(b) vector **e** is a unit vector, i.e. $\mathbf{e}^2 \equiv 1$ and $(\mathbf{e}, \dot{\mathbf{e}}) = 0$;

(c) vector **e** does not rotate in the plane of analysis (the beam cross section).

From condition (b), it follows that $\dot{\mathbf{e}} \perp \mathbf{e}$ and the time derivative $\dot{\mathbf{e}}$ can be represented through a complete set of vectors orthogonal to **e**

$$\dot{\mathbf{e}} = a\mathbf{t} + b[\mathbf{t}, \mathbf{e}].$$

Then, condition (c) imposes constraints for the projection of $\dot{\mathbf{e}}$ onto the beam cross-section plane (normal to $\mathbf{t}$):

$$\dot{\mathbf{e}}_\perp = \dot{\mathbf{e}} - \mathbf{t}(\dot{\mathbf{e}},\mathbf{t}) = b[\mathbf{t},\mathbf{e}] = 0, \qquad (17)$$

whence the equality $b = 0$ follows (by the way, comparison of the first Eq. (15) and Eq. (17) makes it obvious that $\mathbf{V}_\perp(\theta)$ is the velocity of rotation which is seen with respect to $\mathbf{e}$!). At last, with the help of condition (a) we find $a = -(\dot{\mathbf{t}},\mathbf{e})$ and obtain the differential equation

$$\dot{\mathbf{e}} = -(\dot{\mathbf{t}},\mathbf{e})\mathbf{t}. \qquad (18)$$

Equation (18) for components of $\mathbf{e}$ in the laboratory frame is obtained in Appendix 2. It has the form

$$\begin{cases} e_X(\theta) = \cos 2\beta \cos\theta \cos(\theta\cos 2\beta - \alpha) + \sin\theta \sin(\theta\cos 2\beta - \alpha), \\ e_Y(\theta) = \cos 2\beta \sin\theta \cos(\theta\cos 2\beta - \alpha) - \cos\theta \sin(\theta\cos 2\beta - \alpha), \\ e_Z(\theta) = \sin 2\beta \cos(\theta\cos 2\beta - \alpha), \end{cases} \qquad (19)$$

where $\alpha$ is an arbitrary parameter defining the initial position of $\mathbf{e}$:

$$e_X(0) = \cos 2\beta \cos\alpha, \quad e_Y(0) = \sin\alpha, \quad e_Z(0) = \sin 2\beta \cos\alpha.$$

Interestingly, although the velocity of the vector $\mathbf{e}$ rotation in the plane of analysis equals to zero, after the whole cycle of rotation ($\theta = 2\pi$) its position differs from the initial one:

$$e_X(2\pi) = \cos 2\beta \cos(2\pi\cos 2\beta - \alpha), \quad e_Y(2\pi) = -\sin(2\pi\cos 2\beta - \alpha),$$
$$e_Z(2\pi) = \sin 2\beta \cos(2\pi\cos 2\beta - \alpha).$$

This can be seen more clearly in the coordinate system $x'y'z$ ($\theta$), where the beam cross section coincides with the coordinate plane $z = 0$ (this can be done by means of transformation matrix (A.3)). In this frame, components of vector $\mathbf{e}$ amount to

$$\begin{cases} e'_x = e_x \cos\theta - e_y \sin\theta = -\cos(\theta - \theta\cos 2\beta + \alpha), \\ e'_y = e_x \sin\theta + e_y \cos\theta = -\sin(\theta - \theta\cos 2\beta + \alpha), \\ e'_z = 0. \end{cases} \qquad (20)$$

Evidently, in the system $x'y'z$ ($\theta$) vector $\mathbf{e}$ rotates with the velocity

$$\dot{\theta} - \dot{\theta}\cos 2\beta = 2\Omega \sin^2\beta, \qquad (21)$$

which in sum with the "proper" beam rotation velocity (16) gives the total velocity $2\Omega$. Thus, reproducibility of the beam transverse structure after the full revolution of the tilted mirror is ensured. In particular, compare values of $\mathbf{e}$ before and after the whole revolution:

$$e_x(0) = -\cos\alpha, \quad e_y(0) = -\sin\alpha,$$
$$e_x(2\pi) = -\cos(\alpha - 2\pi\cos 2\beta) = -\cos[\alpha + 2\pi(1 - \cos 2\beta)],$$
$$e_y(2\pi) = -\sin(\alpha - 2\pi\cos 2\beta) = -\sin[\alpha + 2\pi(1 - \cos 2\beta)]$$

(see Fig. 8). Therefore, after the whole cycle of the mirror revolution vector **e** (the "natural" reference axis) appears to turn by the angle $2\pi(1-\cos 2\beta) = 2\cdot 2\pi \sin^2\beta$ with respect to its initial position. This is just the quantity which together with the "proper" beam turn (10) amounts to the integer number of cycles ($4\pi$ radians), and the visible transverse pattern of the beam is reproduced.

According to (12) and (16), the velocities of the beam rotation in the coordinate system $x'y'z(\theta)$ and in the laboratory frame differ by $\Omega' - \Omega_l = \Delta\Omega = (1-\cos 2\beta)\Omega$. Our last results imply that the velocity $2\Omega\sin^2\beta$ of vector **e** rotation in the frame $x'y'z(\theta)$ differs from the corresponding velocity measured in the laboratory frame by the same quantity.

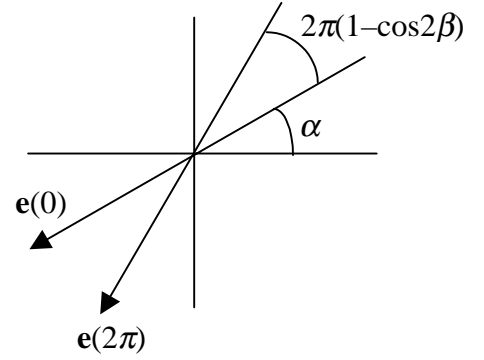

Fig. 8. Reference axis positions (explanations in text)

Note also that, in accord to Eqs. (12) and (21), in the "natural" frame the beam rotates with the velocity

$$2\Omega - 2\Omega\sin^2\beta = 2\Omega\cos^2\beta \qquad (22)$$

and the corresponding angular displacement of the beam transverse pattern $2\theta\cos^2\beta$ changes continuously with the mirror tilt angle $\beta$. In the limit cases of "forward" ($\beta = 0$) and back ($\beta = \pi/2$) reflection the "double" or zero rotation takes place, in full agreement with the real observations. At last, angular velocity (22), visible by the "natural" observer, exactly corresponds to the frequency shift (7) dictated by the energy approach (remember that for the output beam the azimuthal index is $-l$). Therefore, for the "natural" observer whose motion has just been described both paradoxes formulated in Sec. 3 are completely eliminated.

## 5. CONCLUSIONS

The frequency variation of a helical light beam, which is rotated around the axis differing from its own axis (non-collinear RDE) in the arrangement of conical evolution (Fig. 3), has been studied. The examination of the energy exchange between the beam and the evolving optical system has led to the frequency shift (7) that, in contrast to usual practice, is equivalent to the certain beam pattern revolution after the evolving system returns to its initial state (see Eq. (10)). To interpret this conclusion, the especial importance of the observer's motion for the non-collinear RDE has to be realized: Since the beam moves, the observer, whose plane of analysis is always orthogonal to the beam axis, "by the definition" cannot be stationary. The reference axis for the beam angular displacements also cannot be fixed and the question appears which motion of the observer is "natural".

Further study has shown that the non-collinear RDE belongs to a family of impressive and non-trivial (though very clear in concept) phenomena accompanying the process of conical evolution of a light beam (see Fig. 3). Firstly, the frequency shift (7) agrees with angular velocity (16) of the "proper" rotation of the transverse beam pattern around its own axis which is not multiple of the system rotation velocity. Secondly, the visible motion of the beam image projection on a screen changes, seemingly, "by jump" when the mirror tilt angle varies continuously (Fig. 5). Thirdly, the vector within the beam cross section, for which we have postulated zero rotation velocity (**e**), does not return to its initial position after the closed cycle of the beam evolution (see Eq. (20)).

And, the most remarkably, if the vector **e** is chosen as the reference axis, in this frame the observer will see exactly the "proper" rotation velocity (16) and the "energy-dictated" non-collinear RDE frequency shift (7). Besides, for such a choice of the reference axis, the image rotation velocity changes continuously with the mirror tilt angle: Apparent "jumps" in the image behavior between the cases of "forward" and "back" reflection (Fig. 5) appeared because of attempts to fix the observer's position or to change it discontinuously. Therefore, the vector **e** just defines the "natural" observer's frame which is adequate for the conical beam evolution.

The mentioned phenomena have an exclusively geometric nature and are connected to the peculiarities of the superposition of rotational motions around different non-collinear axes. Their analogues occur in different branches of physics, begin-ning with the classical mechanics [18]. For example, Eq. (18) defines the vector **e** parallel translation over a unit sphere surface [19]. In general case, it follows from this equation that after translation along a closed contour, the vector appears to be turned with respect to its initial position [20]. The derivation of Eq. (18) in Sec. 4 almost exactly reproduces the reasoning presented in Ref. [19] for substantiation of the Rytov's law for the polarization plane rotation of a twisted light ray. The motion of the "natural" reference system with respect to the "earthlike" coordinate frame $xyz(\theta)$ expressed by Eq.(19) is identical to the motion of the oscillation plane of the famous Foucault pendulum [18, 20].

From the general point of view, these phenomena are manifestations of the Hannay's geometric phase (a special case of the Berry's topological phase [5, 19, 20]) appearing in the course of adiabatic evolution of non-holonomic systems [21]. The topological nature of the effects makes them insensitive to the specific way of the beam evolution; the analogous non-integer phase shift and the reference system rotation will take place upon the beam evolution along arbitrary closed contour on the unit sphere (the effect value depends only on the solid angle embraced by the contour). These effects are not seen by an "extrincic" (with respect to the beam) observer because the "non-entireness" of the beam revolution is exactly compensated by the angular displacement of the reference system. Nevertheless, their existence is implicitly manifested in the peculiar behavior of the beam pattern rotation depending on the mirror tilt angle (Fig. 5).

Finally, we would like to mention that the analogous conical beam evolution can be realized not only due to reflection (Fig. 3) but with the use of a deflecting system (Fig. 9 in Appendix 3). In this case, the results appear quite similar to those considered above (see Appendix 3); at the same time they reveal a remarkable symmetry and complementarity between the "reflector" and "deflector" schemes of the beam evolution.

## 6. ACKNOWLEDGEMENTS

Authors are grateful to M.S. Soskin and M.V. Vasnetsov for stimulating discussions and for the permission to use Fig. 5 part of which was demonstrated in Ref. [17].

## Appendix 1

The conversion between frames $XYZ$ and $xyz$ or $x'y'z'$ is performed by means of the linear transformations

$$\mathbf{R} = \begin{pmatrix} X \\ Y \\ Z \end{pmatrix} = T(\theta) \begin{pmatrix} x \\ y \\ z \end{pmatrix} = T'(\theta) \begin{pmatrix} x' \\ y' \\ z \end{pmatrix}, \tag{A.1}$$

where

$$T(\theta) = \begin{pmatrix} -\cos 2\beta \cos\theta & \sin\theta & -\sin 2\beta \cos\theta \\ -\cos 2\beta \sin\theta & -\cos\theta & -\sin 2\beta \sin\theta \\ -\sin 2\beta & 0 & \cos 2\beta \end{pmatrix}, \tag{A.2}$$

$$T'(\theta) = \begin{pmatrix} -(\cos^2\beta - \sin^2\beta \cos 2\theta) & \sin^2\beta \sin 2\theta & -\sin 2\beta \cos\theta \\ \sin^2\beta \sin 2\theta & -(\cos^2\beta + \sin^2\beta \cos 2\theta) & -\sin 2\beta \sin\theta \\ -\sin 2\beta \cos\theta & -\sin 2\beta \sin\theta & \cos 2\beta \end{pmatrix}. \tag{A.3}$$

Note that this transformation matrix is symmetric and, therefore, is valid both for the direct and inverse transform expressed by the second equation (A.1).

## Appendix 2

Expressing $\mathbf{e}$ by means of Cartesian components and making use of Eq. (11) and equality $\dot{\mathbf{e}} = (d\mathbf{e}/d\theta)\dot{\theta}$, we obtain from Eq. (18)

$$\frac{de_X}{d\theta} = -v \sin^2 2\beta \cos\theta, \quad \frac{de_Y}{d\theta} = -v \sin^2 2\beta \sin\theta, \tag{A.4}$$

where

$$v = -e_X \sin\theta + e_Y \cos\theta. \tag{A.5}$$

In the similar manner one can write down an equation for the third component but it is not necessary because the condition of orthogonality $(\mathbf{e}, \mathbf{t}) \equiv 0$ allows to obtain

$$e_Z = u \tan 2\beta, \tag{A.6}$$

where
$$u = e_X \cos\theta + e_Y \sin\theta. \qquad (A.7)$$

Introducing the new variables (A.5) and (A.7) transforms equations (A.4) to the easily solvable form
$$\frac{du}{d\theta} = v\cos^2 2\beta, \quad \frac{dv}{d\theta} = -u.$$

Taking into account condition (b) of Sec. 4, i.e. $u^2 + v^2 + e_Z^2 = e_X^2 + e_Y^2 + e_Z^2 = 1$, and using (A.5) – (A.7) leads to (19).

## Appendix 3

In order to show that our results relate not only to the chosen variant of performing the beam evolution, consider the situation where the beam deviates by the same angle $2\beta$ with a deflector. In Fig. 5, there is presented the arrangement with the symmetric prism that can move around the axis $Z$ (the beam axis follows a polyline experiencing the double refraction on the prism facets and point $M$, as before, lies at the intersection of the output beam axis and axis $Z$).

The main distinction of this case is that the beam transverse pattern experiences no inversion upon the "forward" deflection. As before, the unit vector of the moving axis is determined by Eq. (11). The image of the axis $X$ unit vector, similarly to (13), obeys the equation

$$\mathbf{R}(\theta) = \begin{pmatrix} \sin^2\beta - \cos^2\beta \cos 2\theta \\ -\cos^2\beta \sin 2\theta \\ -\sin 2\beta \cos\theta \end{pmatrix}.$$

Operating as before, we find the velocity of the end of this vector (analogue of Eq. (14))

$$\mathbf{V}(\theta) = \dot{\mathbf{R}}(\theta) = \Omega \begin{pmatrix} 2\sin^2\beta \sin 2\theta \\ -2\sin^2\beta \cos 2\theta \\ -\sin 2\beta \sin\theta \end{pmatrix},$$

its component within the beam cross-section and corresponding angular velocity (see (15), (16))

$$\mathbf{V}_\perp(\theta) = 2\Omega \sin^2\beta \begin{pmatrix} \sin^2\beta \sin 2\theta \\ -\cos^2\beta - \sin^2\beta \cos 2\theta \\ -\sin 2\beta \sin\theta \end{pmatrix},$$

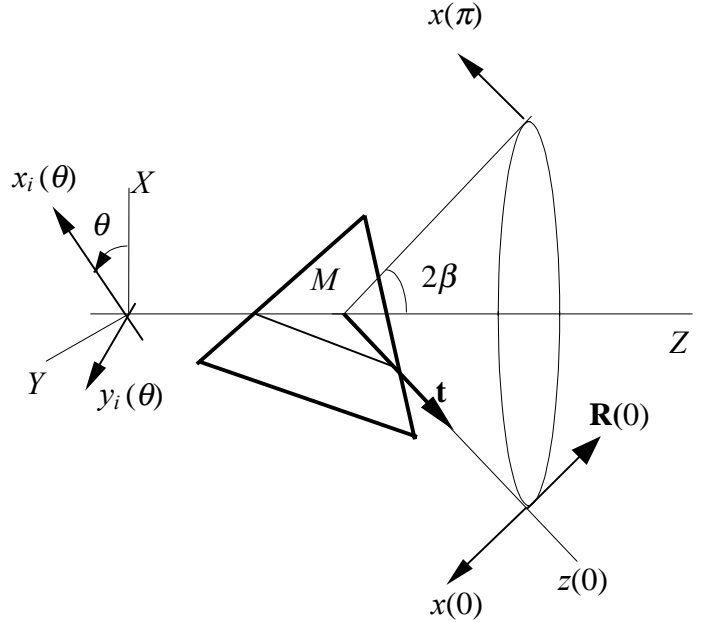

Fig. 9. A deflector scheme for the beam conical evolution (compare to Fig. 7 that can be called a "reflector scheme").

$$\mathbf{\Omega}_1 = \left[\mathbf{R}(\Theta), \mathbf{V}_\perp(\theta)\right] = -2\Omega \sin^2\beta \cdot \mathbf{t}.$$

Interestingly, here the "proper" beam rotation and the deflector rotation have opposite directions. Like in the case of rotating mirror, the reference axis can be associated with the same vector **e** (21). Taking its motion into account permits to describe the output beam rotation in the "natural" frame. Note that, as for the image rotation, the case of β < 45° ("forward" deflection) is analogous to the situation of β > 45° ("back" reflection) in the mirror scheme (see Figs. 5–7), and vice versa.